\documentclass{emulateapj}
\usepackage{amsmath}
\usepackage{amssymb}
\usepackage{graphicx}

\newcommand{\mytilde}{\raise.17ex\hbox{$\scriptstyle\mathtt{\sim}$}}

\begin{document}
\title{K2 Variable Catalogue I: A catalogue of Variable Stars from K2 Field 0}

\author{D. J. Armstrong, H. P. Osborn, D. J. A. Brown, J. Kirk, K. W. F. Lam, D. L. Pollacco, J. Spake, S. R. Walker}
\affil{University of Warwick, UK}
\email{d.j.armstrong@warwick.ac.uk}

 \begin{abstract}
We have searched the K2 campaign 0 data for lightcurve variations associated with stellar variability. The results of this search are presented as a catalogue, giving the identifiers of nearly 2500 variable stars in the dataset. We list the detected range of the variation, periodicity if relevant and semi-amplitude. Lightcurves are classified into strictly periodic, quasi-periodic and aperiodic groups. We do not attempt to identify the source of variability, which may arise from pulsation or stellar activity. However, we cross match the objects against variable star related guest observer proposals, specifying the variable type in many cases. At present eclipsing binary stars are not included. Future releases will address each K2 field as it is made available, and may be improved to include more detailed catalogue information and to provide detrended object lightcurves.
 \end{abstract}

\section{Introduction}
The K2 mission \citep{Howell:2014ju} is the next generation of the Kepler space telescope, and became fully operational in June 2014. It will survey a series of fields near the ecliptic, returning continuous high-precision data over an 80 day period for each field. Despite the reaction wheel losses that stopped the Kepler mission, K2 has been estimated to be capable of 50ppm precision, close to the sensitivity of the primary mission. All data will be public, although so far only campaign 0 has been released, in September 2014. As the mission progresses, much more data should become available, with campaign 1 expected in November 2014. Targets are provided by the Ecliptic Plane Input Catalogue (EPIC) which is hosted at the Mikulski Archive for Space Telescopes (MAST) along with the available data products. Approximately 7500 objects were observed during campaign 0, mostly in `long-cadence' (a cadence of \mytilde 30 min). A few (13) were also observed in `short-cadence' (\mytilde 1 min). All identification processes in this catalogue were performed on the long cadence dataset.

The K2 mission will be of great use to a wide range of astronomical research areas. Although the original Kepler space telescope was primarily aimed at the detection and study of exoplanets, its high precision lightcurves were used for studies with astroseismology \citep[e.g.][]{Chaplin:2013jf}, stellar rotation \citep[e.g.][]{Reinhold:2013iz} and eclipsing binaries \citep[e.g.][]{Prsa:2011dx}, to name just a few. Already the K2 mission has been used to identify new candidate eclipsing binaries \citep{Conroy:2014gy}. The utility of Kepler extended to the study of variable stars, with a number of studies en masse and individually of different kinds of variable \citep[e.g.][]{McQuillan:2012dj,Holdsworth:2014hc,Stello:2014is,Banyai:2013hu}. Catalogues were made available using a variety of techniques \citep{Debosscher:2011kz,Uytterhoeven:2011jv}. Recently such catalogues have begun appearing for the K2 mission, including a recent cross match with the TESS target catalogue \citep{Stassun:2014wz}. Here we present a further catalogue for the K2 data, identifying and classifying stars showing variability in the first field of K2 observations. We hope that this catalogue will prove useful to both variable star studies and other users of the K2 data. Future releases will include more K2 fields, as well as other planned improvements. We are open to suggestions by the community as to what new improvements to implement.

\begin{figure}
\resizebox{\hsize}{!}{\includegraphics{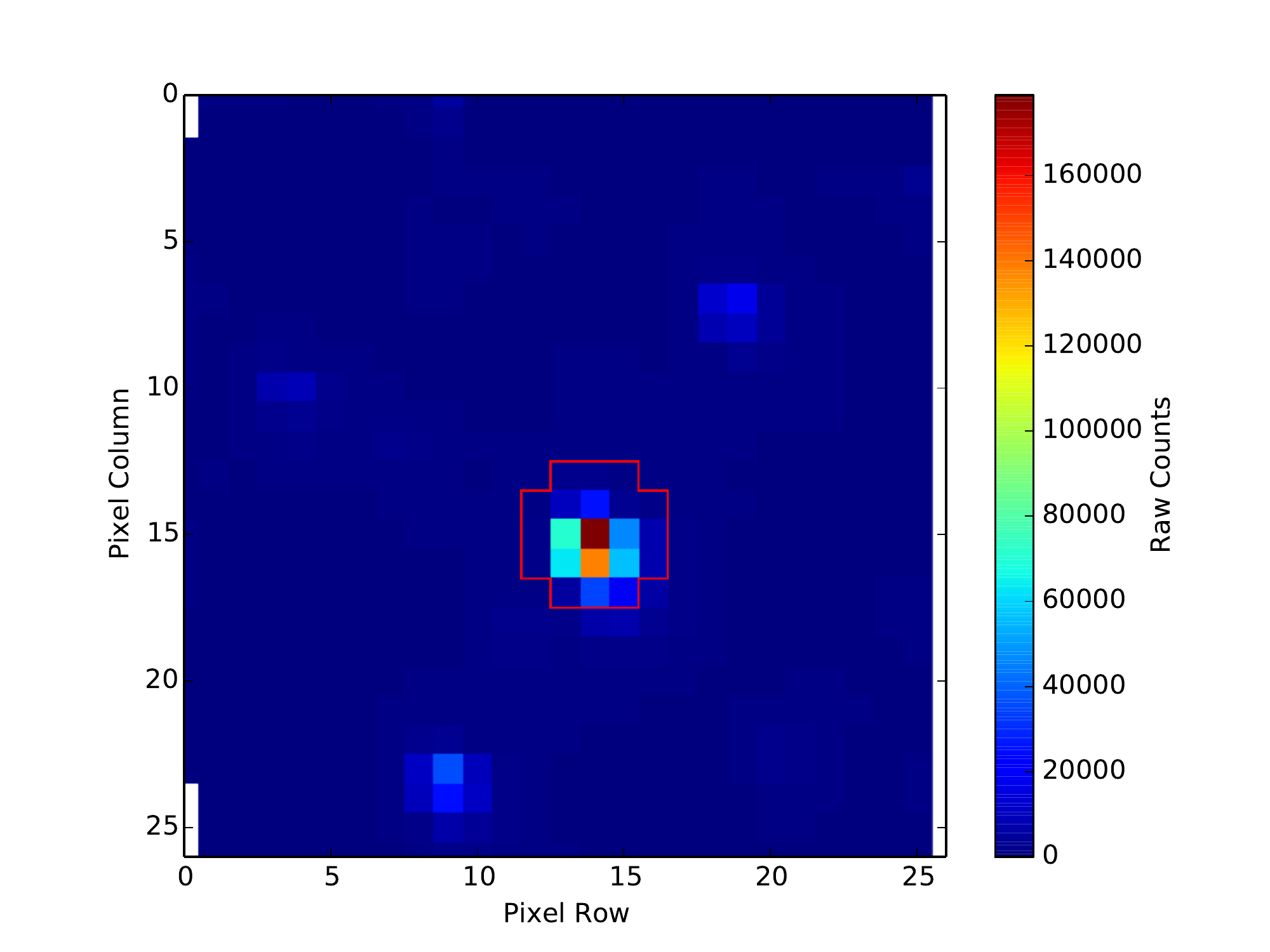}}
\caption{Example target pixel file and aperture. In this case a single time slice from EPIC202072369 is shown. The aperture is represented by the red solid line.}
\label{figextract}
\end{figure}

\begin{figure*}
\resizebox{\hsize}{!}{\includegraphics{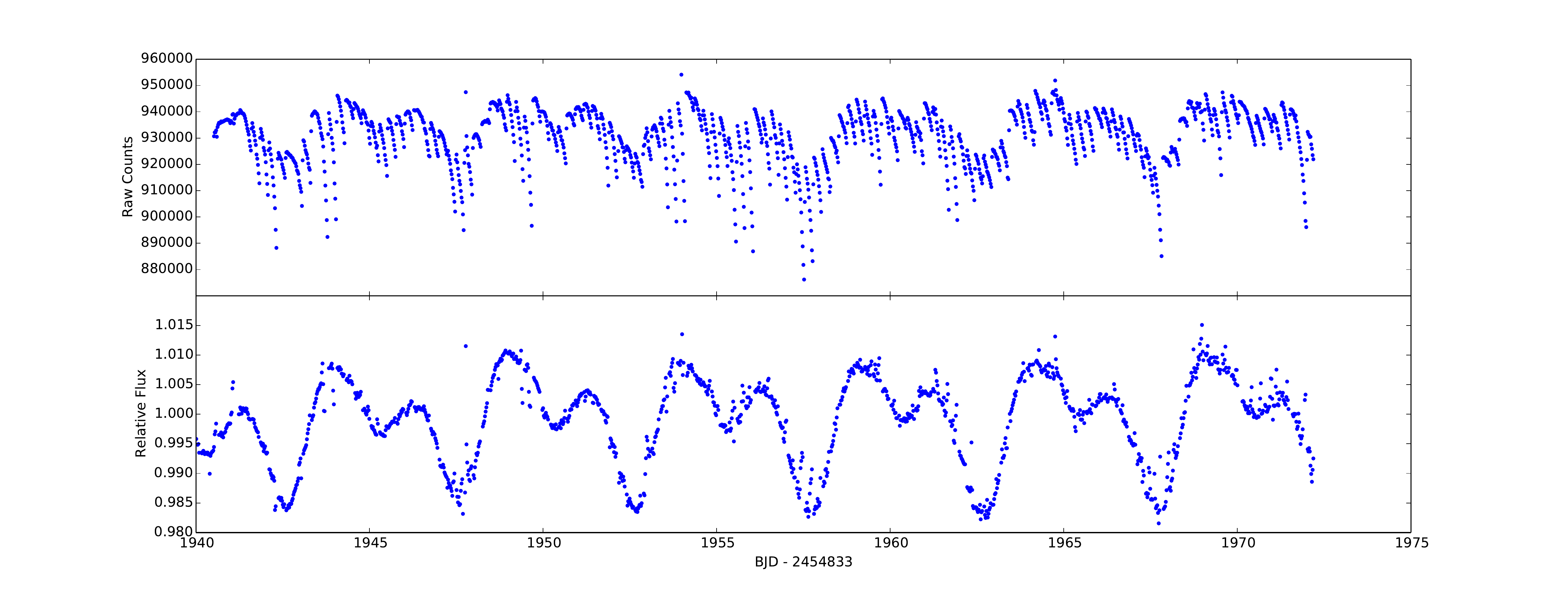}}
\caption{The lightcurve of EPIC202059229 (a QP classified object) before (above) and after (below) detrending.}
\label{figdetrend}
\end{figure*}

\section{Data Preparation}
\subsection{Data Source}
Our lightcurves were obtained from the MAST archive of K2 data (Data Release 1). These are at present only available as Target Pixel Files, giving the pixel time series of a variably sized window surrounding the proposed target. At this stage we used only the long cadence observations (bearing in mind that each short cadence target also has data in long cadence). To make use of these files, we needed to perform photometry on the window. Initially the data were cut to give only times later than 1940.5 (BJD-2454833), leaving a baseline of 32 days. This was due to coarse point and safe mode events which occurred through the first half of the K2 campaign 0 dataset. For this catalogue release the pixel window was cropped to a centrally placed 9x9 box. The brightest pixel within this box on average over the time series was then found and an aperture placed, centred on this pixel. Data Release 2 of campaign 0 was released late in the preparation of this catalogue --- this later release provided useable wcs information for the target pixel files, and later releases of this catalogue will make use of this for improved target identification. The aperture shape was identical for each target. See Figure \ref{figextract} for an example pixel window and aperture. Once a raw lightcurve was available, background subtraction was performed in the usual way.

We warn users that for faint targets with brighter companions within \mytilde20" there is a significant chance that the bright companion was extracted rather than the proposed target. This makes the information available within the EPIC catalogue erroneous (in that it applies to a different object than the one in this catalogue), and caution should be advised. 

\subsection{Data Detrending}

The main source of systematic noise in K2 data is pointing drift, as has been pointed out previously \citep{Vanderburg:2014bi}. We independently implemented a method similar to that proposed by \citet{Vanderburg:2014bi} in order to detrend our lightcurves. Initially the row and column centroid positions were calculated for each timestamp. At this stage points near a thruster firing event were cut, detected as those to either side of times where the point-to-point centroid shift was greater than 3 times the median point-to-point shift across the dataset. The centroid positions were then used to create a 2D surface of raw flux against position of the centroid on the CCD. If the pointing drift had no impact on the flux, this surface should show no correlation. Instead, in the majority of cases a strong trend was seen. This trend was identified through binning the data into 10 evenly spaced bins in row and 10 in column. The median flux in each bin was then taken and interpolated between linearly, creating a smooth surface mapping the variation caused by the observed centroid shifts. This surface was divided out, effectively decorrelating the flux from spacecraft pointing and providing a lightcurve in relative flux. In addition, outliers were removed by cutting data points where the centroid position was greater than 5 times the median distance from the median centroid position across the dataset. In all these situations medians rather than means and standard deviations were used in order to avoid the effects of large outliers. An example lightcurve, pre and post detrend, is shown in Figure \ref{figdetrend}.

We found that this method worked well in most cases. Its weakness lies in longer period variability - intrinsic stellar variability which occurs on a similar timescale to the dataset can be removed, if the spacecraft drift spuriously correlates with it. As such variability on timescales longer than 10 days, while in the catalogue, may be affected. Large amplitude variability which dominates over the pointing noise can also be reduced in amplitude, should it correlate with the drift. We note however that as a beneficial side effect this method of detrending automatically weakens signals associated with variability on a background blended object, as such variability can cause centroid position changes correlated with the change in flux. This applies equally to stellar variability or to background blended eclipses.

\section{Catalogue}
\begin{table*}
\centering
\caption{Catalogue Table. An extract from the table is shown. The full table is available online.}
\label{tabcatalogue}

\begin{tabular}{@{}lllllr@{}}
\hline
EPIC ID & Type & Range & Period & Amplitude &Proposal Information\\
         &    & \%  &   d & \% & \\
\hline
\hline
:&:&:&:&: &: \\
202059697& P&4.42&3.423396&1.89&\\
202060097& P&2.23&2.624037&0.85&\\
202060098&QP&1.63&4.989129&0.42&\\
202060130&AP&0.86&0.000000&0.00&49 (SpB star)\\
202060132&QP&7.96&0.734701&3.59&44 (Beta Ceph);49 (SpB star)\\
:&:&:&:&: &: \\
\hline

\end{tabular}

\end{table*}

The catalogue is presented in Table \ref{tabcatalogue}, and the full version can be found online\footnote{http://deneb.astro.warwick.ac.uk/phrlbj/k2varcat/}. For each lightcurve a Lomb-Scargle periodogram was created \citep{Lomb:1976bo,Scargle:1982eu} then using this and the detrended flux the object was classified by eye. Lightcurves with no variability showing above the noise were rejected. We note that this noise threshold can change significantly for individual stars - as such, the absence of an object from this catalogue does not imply that it is less variable than any object in the catalogue, only that any variability it has was hidden behind either white or remnant systematic noise. At present, eclipsing binary objects are not included.

\subsection{Fields}
\begin{enumerate}
\item \textbf{EPIC ID}\\
ID of target from the EPIC catalogue. Spans 202059074--202139979.
\item \textbf{Type}\\
 Lightcurves were classified by eye as Periodic (P), Quasi-Periodic (QP), or Aperiodic (AP). Periodic classification implies a sinusoidal variation of constant period and amplitude. Quasi-periodic objects have amplitude or period variations, or a lightcurve non-sinusoidal in shape. Aperiodic objects showed no periodicity. In many cases these objects may be periodic but with periods greater than \mytilde15 days, a limit imposed by the data baseline. Users should be aware that objects which should be classified as P can be misclassified as QP due to noise.
\item \textbf{Range}\\ 
The lightcurves were binned into 10 point wide bins and the median of each bin found. The range given is the maximum bin less the minimum, in relative flux units. In some cases outliers or remnant noise can affect this calculation, leading to ranges larger than are shown. Spans 0.035--13.97\%.
\item \textbf{Period}\\
The most significant peak from a Lomb-Scargle periodogram, for objects classified as P or QP. Where possible the true period rather than aliases is given, even if the aliases were more significant. Zero for AP objects. No periods larger than 15 days are shown to avoid spurious detections due to the data baseline (32 days). For the same reason, while we report periods between 10 -- 15 days these should be treated with some caution. Spans 0--14.9579 days.
\item \textbf{Amplitude}\\
The semi-amplitude of the lightcurve at the stated period, for objects classified P or QP. This was calculated through phase-folding the lightcurve, binning it into 40 evenly spaced bins, then taking the median of each bin. The semi-amplitude represents half of the maximum minus minimum bin value, in relative flux units. Short period objects will show reduced amplitude due to the cadence of the observations. Zero for AP objects. Spans 0--32.71\%.
\item \textbf{Proposal Information}\\
Guest Observer proposals relating to the object. Only variable star related proposals are shown. If possible, the specific variable types which each proposal is related to are given in brackets.
\end{enumerate}



\section{Conclusion}
We have presented a catalogue of stars showing variability in the K2 campaign 0 dataset, along with periods and amplitudes. Some of these variables are previously known and result from proposals targeting variable stars, while others are not. In both cases the K2 data will often represent the highest precision photometry ever obtained of the target. We hope that this catalogue will be useful to those studying variable stars, and to those searching for planets in the K2 data, where stellar variability represents a barrier to detection. Future releases of this catalogue will be made as K2 releases more data. These releases will utilise the recently released world coordinate system information for improved target extraction, as well as providing extracted and detrended lightcurves directly from the catalogue. We plan to introduce more categories, improving variable typing. Suggestions from the community as to further improvements are welcome.

\section*{Acknowledgements}
The data presented in this paper were obtained from the Mikulski Archive for Space Telescopes (MAST). STScI is operated by the Association of Universities for Research in Astronomy, Inc., under NASA contract NAS5-26555. Support for MAST for non-HST data is provided by the NASA Office of Space Science via grant NNX13AC07G and by other grants and contracts.

\bibliography{papers201114}

\begin{thebibliography}{}

\bibitem[\protect\citeauthoryear{Banyai et~al.,}{Banyai
  et~al.}{2013}]{Banyai:2013hu}
Banyai E.  et~al., 2013, Monthly Notices of the Royal Astronomical Society,
  436, 1576

\bibitem[\protect\citeauthoryear{Chaplin et~al.,}{Chaplin
  et~al.}{2013}]{Chaplin:2013jf}
Chaplin W.~J.  et~al., 2013, The Astrophysical Journal Supplement Series, 210,
  1

\bibitem[\protect\citeauthoryear{Conroy et~al.,}{Conroy
  et~al.}{2014}]{Conroy:2014gy}
Conroy K.~E.  et~al., 2014, Publications of the Astronomical Society of the
  Pacific, 126, 914

\bibitem[\protect\citeauthoryear{Debosscher, Blomme, Aerts \&
  De~Ridder}{Debosscher et~al.}{2011}]{Debosscher:2011kz}
Debosscher J.,  Blomme J.,  Aerts C.,    De~Ridder J.,  2011, Astronomy and
  Astrophysics, 529, A89

\bibitem[\protect\citeauthoryear{Holdsworth, Smalley, Kurtz, Southworth, Cunha
  \& Clubb}{Holdsworth et~al.}{2014}]{Holdsworth:2014hc}
Holdsworth D.~L.,  Smalley B.,  Kurtz D.~W.,  Southworth J.,  Cunha M.~S.,
  Clubb K.~I.,  2014, Monthly Notices of the Royal Astronomical Society, 443,
  2049

\bibitem[\protect\citeauthoryear{Howell et~al.,}{Howell
  et~al.}{2014}]{Howell:2014ju}
Howell S.~B.  et~al., 2014, Publications of the Astronomical Society of the
  Pacific, 126, 398

\bibitem[\protect\citeauthoryear{Lomb}{Lomb}{1976}]{Lomb:1976bo}
Lomb N.~R.,  1976, Astrophysics and Space Science, 39, 447

\bibitem[\protect\citeauthoryear{McQuillan, Aigrain \& Roberts}{McQuillan
  et~al.}{2012}]{McQuillan:2012dj}
McQuillan A.,  Aigrain S.,    Roberts S.,  2012, Astronomy and Astrophysics,
  539, A137

\bibitem[\protect\citeauthoryear{Prsa et~al.,}{Prsa et~al.}{2011}]{Prsa:2011dx}
Prsa A.  et~al., 2011, The Astronomical Journal, 141, 83

\bibitem[\protect\citeauthoryear{Reinhold, Reiners \& Basri}{Reinhold
  et~al.}{2013}]{Reinhold:2013iz}
Reinhold T.,  Reiners A.,    Basri G.,  2013, Astronomy and Astrophysics, 560,
  A4

\bibitem[\protect\citeauthoryear{Scargle}{Scargle}{1982}]{Scargle:1982eu}
Scargle J.~D.,  1982, The Astrophysical Journal, 263, 835

\bibitem[\protect\citeauthoryear{Stassun, Pepper, Paegert, De~Lee \&
  Sanchis-Ojeda}{Stassun et~al.}{2014}]{Stassun:2014wz}
Stassun K.~G.,  Pepper J.~A.,  Paegert M.,  De~Lee N.,    Sanchis-Ojeda R.,
  2014, eprint arXiv:1410.6379

\bibitem[\protect\citeauthoryear{Stello et~al.,}{Stello
  et~al.}{2014}]{Stello:2014is}
Stello D.  et~al., 2014, The Astrophysical Journal, 788, L10

\bibitem[\protect\citeauthoryear{Uytterhoeven et~al.,}{Uytterhoeven
  et~al.}{2011}]{Uytterhoeven:2011jv}
Uytterhoeven K.  et~al., 2011, Astronomy and Astrophysics, 534, A125

\bibitem[\protect\citeauthoryear{Vanderburg \& Johnson}{Vanderburg \&
  Johnson}{2014}]{Vanderburg:2014bi}
Vanderburg A.,  Johnson J.~A.,  2014, Publications of the Astronomical Society
  of the Pacific, 126, 948

\end{thebibliography}
\bibliographystyle{mn2e_fix}

\end{document}